\documentclass[10pt]{scrartcl}
\usepackage[margin=1.25cm]{geometry}
\usepackage{graphicx}
\usepackage{dcolumn}
\usepackage{bm}

\usepackage{stackengine}

\stackMath
\usepackage{svg}
\usepackage{pdfpages}
\usepackage{array}
\usepackage{multirow}
\usepackage{amssymb}
\usepackage{amsmath}
\usepackage{import}
\usepackage{graphicx}
\usepackage{color}
\usepackage{blindtext}
\usepackage{subcaption}
\usepackage{physics}
\usepackage{aligned-overset}
\usepackage{placeins}

\usepackage[utf8]{inputenc}
\usepackage[T1]{fontenc}
\usepackage{mathptmx}
\usepackage{mathrsfs}  
\DeclareMathAlphabet{\mathcal}{OMS}{cmsy}{m}{n}

\usepackage{hyperref}

\begin{document}
\newcommand{\wavtrans}{T}
\def\printImages{}  

\renewcommand{\vec}[1]{\textbf{\textit{#1}}}
\renewcommand{\rm}[1]{\text{#1}}

\title{Fast Simulation of Wide-Angle Coherent Diffractive Imaging}
\date{}
\maketitle
\noindent
P. Tuemmler, J. Apportin, T. Fennel*, C. Peltz*\\
University of Rostock, Institute of Physics\\
Albert-Einstein-Str. 23 18059 Rostock, Germany\\
thomas.fennel@uni-rostock.de, christian.peltz@uni-rostock.de

\begin{abstract}
\noindent
Single-shot coherent diffractive imaging (CDI) using intense XUV and soft X-ray pulses holds the promise to deliver information on the three dimensional shape as well as the optical properties of nano-scale objects in a single diffraction image. This advantage over conventional X-ray diffraction methods comes at the cost of a much more complex description of the underlying scattering process due to the importance of wide-angle scattering and propagation effects.
The commonly employed reconstruction of the sample properties via iterative forward fitting of diffraction patterns requires an accurate and fast method to simulate the scattering process. 
This work introduces the propagation multi-slice Fourier transform method (pMSFT) and demonstrates its superior performance and accuracy against existing methods for wide-angle scattering. A derivation from first principles, a unified physical picture of the approximations underlying pMSFT and the existing methods, as well as a systematic benchmark that provides qualified guidance for the selection of the appropriate scattering method is presented.
\end{abstract}

\section{Introduction}

Single-shot coherent diffractive imaging (CDI) has matured into a well-established tool for the characterization of the structure and dynamics of nanoparticles in free flight~\cite{Miao1999,Chapman2010,Miao2012}. Applications ranging from the morphology analysis of aerosols~\cite{Loh2012}, over the visualization of vortices in superfluid helium droplets~\cite{Langbehn2018, Gomez2014, Langbehn2022}, to the characterization of the shape and ultrafast structural dynamics of atomic clusters~\cite{Rupp2012, Barke2015, Gorkhover2018, Peltz2022, Bacellar2022, Shin2023, Dold2025} and biological samples~\cite{Seibert2011, Ekeberg2015, Ekeberg2024} highlight the remarkable versatility of the method. While single-shot CDI has become feasible first in the X-ray domain~\cite{Seibert2011} thanks to the enormous advances of X-ray free electron lasers in the last two decades~\cite{Ackermann2007, Schoenlein2019}, the method has meanwhile expanded into the XUV and soft X-ray range~\cite{Barke2015, Langbehn2018} that is also accessible for lab based experiments using intense high harmonic generation sources~\cite{Rupp2017}.

The major challenge in analyzing CDI experiments is the extraction of the sample properties encoded in the measured diffraction pattern. In the X-ray regime, shape information is encoded in the signal under small scattering angles and the projected density of the target can be extracted using phase retrieval algorithms~\cite{Fienup1982, Martin2012, Shechtman2015, Colombo2025}. The underlying approximation that the scattering can be described by a single two-dimensional projection of the scattering strength, which is commonly interpreted as a density projection along the optical axis, breaks down as soon as wide-angle scattering as well as propagation effects contribute significantly, which is typically the case in the XUV and soft X-ray range. On the one hand, the description of the scattering process becomes much more challenging in  this case and in principle requires the full solution of Maxwell's equations, e.g. via Finite-difference Time-Domain methods (FDTD). On the other hand, the wide-angle scattering features and the specific material sensitivity emerging in the XUV and soft X-ray range provide valuable information on both the three-dimensional shape~\cite{Barke2015} as well as frequency-dependent optical properties of the target~\cite{Rupp2017} in a single shot scattering image.

Unfortunately, no rigorous inversion methods are known so far to directly extract this information from the measured scattering pattern. To date, the most common reconstruction approach is iterative forward fitting of the measured scattering images by optimizing the parameters of a suitable geometry model~\cite{Barke2015, Langbehn2018, Rupp2017, Colombo2023}. The practical application of this strategy requires a sufficiently fast and accurate method to simulate the scattering process for a large number of geometry candidates, effectively ruling out fully fledged approaches like FDTD~\cite{Barke2015} and DDA~\cite{Sander2015}. 

The most commonly applied approximate method for the simulation of wide-angle scattering pattern for the geometry candidates is the multi-slice Fourier transform (MSFT) method~\cite{Barke2015, Langbehn2018, Colombo2023}, which yields a speed-up of several orders of magnitude, at the cost of a strongly limited description of propagation effects.

Here, we introduce the propagation multi-slice Fourier transform method (pMSFT) that offers a substantial accuracy improvement over existing wide-angle scattering methods at the same numerical complexity. We derive a split-step representation of pMSFT from first principles starting from the scalar wave equation, provide a slice-resolved notation that reveals the connection to established methods, and present a systematic performance benchmark for a large set of experimentally relevant scenarios. The resulting unified picture of the physics combined with the performance analysis provides qualified guidance on the range of applicability and suitability of the investigated methods, and shows that pMSFT is the method of choice in most cases.

To illustrate the performance of pMSFT in relation to established scattering methods, i.e. Hare's split-step\\method~\cite{Hare1994}, MSFT, the solution in Born's approximation~\cite{Born1926}, and the small-angle X-ray scattering method (SAXS)~\cite{Guinier1979}, we compare their predictions for a representative experimental scenario. The corresponding scattering pattern for a truncated silver octahedron illuminated with 90 eV soft X-ray radiation \cite{Barke2015} are displayed in \textbf{Figure~\ref{fig:fdtd_vs_pmsft}} in terms of the differential scattering cross sections. Comparison with a quasi-exact FDTD reference calculation shows nearly perfect agreement for the case of pMSFT and the successive increase of deviations for the other methods, revealing the substantial impact of the underlying additional approximations. In particular, the prediction of Hare's split-step method substantially deviates at large angles, leading to a misrepresentation of the finger-like streak features. The MSFT method still captures the qualitative structure of the scattering image but shows distortions and quantitative deviations at all angles. The prediction in Born approximation only captures the basic features, such as the three-fold symmetry of the streaks, but misrepresents both, the fringe contrast and the absolute signal strength. As expected for the considered wide-angle test scenario, SAXS performs poorly and only captures the six-fold symmetry in the innermost part of the scattering image, missing essentially all the other features. A quantitative comparison is displayed via the line-outs presented in Figure~\ref{fig:fdtd_vs_pmsft}~\hyperref[fig:fdtd_vs_pmsft]{(b)}.

\ifdefined\printImages
\begin{figure*}
    \fontsize{8}{10}\selectfont
    \centering
    \includegraphics[width=1\hsize]{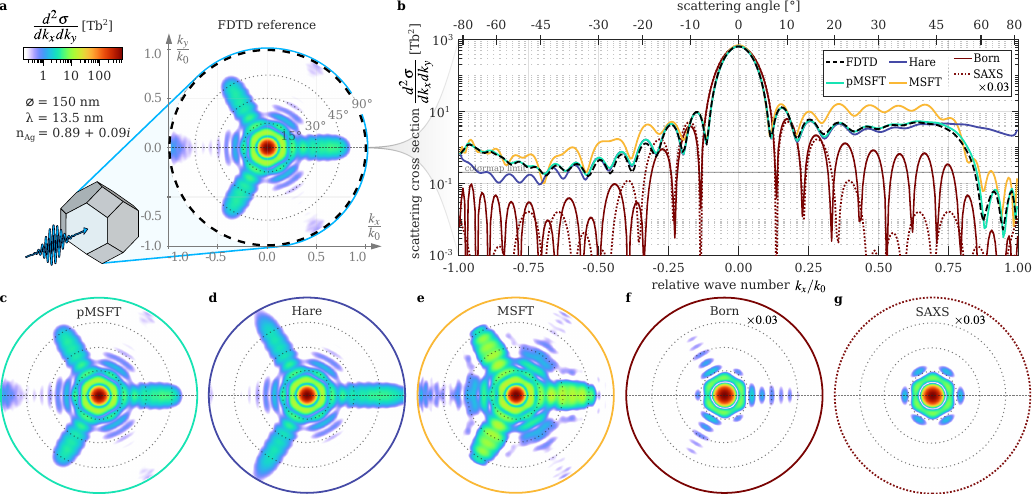}
    \caption{Comparison of the differential scattering cross section of a truncated silver octahedron (150 nm diameter) for illumination with 90 eV (13.5 nm) soft X-ray radiation~\cite{Barke2015}
    as predicted from different scattering simulation methods (as indicated). The result from an FDTD simulation (a) using optical properties of bulk-silver~\cite{Henke1993}  serves as a reference to benchmark the remaining  approaches (c-g), see text for details. 
    The results from pMSFT, Hare's method, MSFT, Born, and SAXS reflect the impact of the increasing level of approximation underlying the specific methods. 
    Horizontal cuts through the respective two-dimensional k-space representations for $k_y=0$ enable the detailed comparison of quantitative signal strength and feature reproduction (b). Note that the Born and SAXS results are down-scaled by a factor of $0.03$ for convenience in all plots and that the data shown in (c-g) includes the polarization correction, see Section \ref{sec:ff_detector}. }
    \label{fig:fdtd_vs_pmsft}
\end{figure*}
\fi 

The remainder of this article is structured as follows. We first present the formal derivation of the pMSFT method from first principles and highlight its connection to the other established methods in Sections \ref{sec:pMSFTderivation} and \ref{sec:pMSFTconnection}. The near to far-field transformation of the predicted exit fields including a polarization correction is described in Section \ref{sec:ff_detector} for different detector geometries. A systematic benchmark of the feature reproduction capabilities and the quantitative signal strength prediction is presented for a large range of optical parameters in Section \ref{sec:benchmark}. A conclusion including a general recommendation for the choice of the scattering simulation method is provided in Section \ref{sec:conclusion}. Note that a set of helpful but lengthy derivations are given in the supplement. 

\section{Derivation of the Propagation Multi-Slice-Fourier-Transform method \label{sec:pMSFTderivation}}
The two major steps in the calculation of scattering images are the propagation of the incident field through the target and the subsequent free-space propagation of the resulting exit field to the detector.  In this section we address the much more challenging first task and present a concise derivation of the pMSFT method. 

Therefore, we consider a linearly polarized monochromatic incident field with vacuum wave number $k_0$ to hit, propagate through, and scatter off a non-magnetic target with a linear isotropic local optical response described by the susceptibility  map $\chi(\mathbf r)$ that also encodes the object's shape. The associated complex electric field $E(\vec r)$ is defined by the scalar Helmholtz equation
\begin{equation}
\left[ \Delta + k_0^2\left(1+\chi(\vec r)\right) \right] E(\vec r)=0,\label{eq:prop_formally}
\end{equation}
see the supporting information for a brief derivation and a justification of the  scalar approximation. For homogeneous media (constant $\chi$) Equation~\eqref{eq:prop_formally} is solved by plane waves $E_{\vec k}(\vec r)=\text{exp}(i \vec k \cdot \vec r)$ with wave number $k=n\,k_0$ given by the refractive index $n=\sqrt{1+\chi}$, which applies, e.g., to the propagation of the exit field to the detector.

In the following, we derive a propagator-based method to solve the above Helmholtz equation. This approach relies on two-dimensional slices of the field distribution and describes their propagation along the optical axis in finite steps. For many applications an incoming field in the form of a plane wave propagating along the optical axis will be of interest, but in general  any incoming field structure, e.g. a focused beam, can be considered. We define the z-axis as the optical axis such that a cut through a scalar field $f(x,y,z)$ at a fixed position $z$ defines the associated two-dimensional slice $f(x,y;z)$ in real space. This is linked to the corresponding transverse Fourier representation $\tilde{f}(k_x,k_y;z)$ in $k$-space by
\begin{align}
    f(z)&=f(x,y;z)=\mathcal{\hat F}^{-1} \tilde{f}(k_x,k_y;z)=\frac{1}{2\pi}\iint_{-\infty}^{\infty}\tilde{f}(k_x,k_y;z)e^{i(k_xx+k_yy)}dk_x dk_y,\quad\text{and}\label{eq:inverse_ft}\\
    \tilde{f}(z)&=\tilde{f}(k_x,k_y;z)=\mathcal{\hat F} f(x,y;z) =\frac{1}{2\pi}\iint_{-\infty}^{\infty}f(x,y;z)e^{-i(k_xx+k_yy)}dxdy,\label{eq:ft}
\end{align}
where the linear operators $\mathcal{\hat F}^{-1}$ and $\mathcal{\hat F}$ represent short-hand notations for the Fourier transforms with $\mathcal{\hat F}^{-1} \mathcal{\hat F}=\hat 1$.

After transforming Equation~\eqref{eq:prop_formally} to transverse $k$-space, one obtains 
\begin{align}
    \frac{\partial^2}{\partial z^2}\tilde{E}(z)=-\biggl[k_z^2+k_0^2\underbrace{\mathcal{\hat F}\chi(z)\mathcal{\hat F}^{-1}}_{\hat M(z)}\biggr]\tilde{E}(z),\label{eq:prop_fourierspace_eq}
\end{align}
where $k_z^2(k_x,k_y)={k_0^2 - k_x^2 - k_y^2}$ and $\chi(z)=\chi(x,y;z)$ represent the squared axial wave number  and the susceptibility map at axial position $z$, respectively.
The first term in the square brackets on the right hand side of Equation~\eqref{eq:prop_fourierspace_eq} results in a trivial product of $k_z^2$ and the field  and represents vacuum propagation. The second term is non-trivial, describes the medium effect associated with the susceptibility, and can be expressed by the medium  operator $\hat{\tilde{M}}(z)$. Note that the application of $\hat{\tilde{M}}(z)$ on the electric field corresponds to a convolution (denoted by $\circledast$) with the susceptibility in k-space
\begin{align}
\hat{\tilde{M}}(z)\tilde{E}(z) = \mathcal{\hat F}\chi(z)\mathcal{\hat F}^{-1}\tilde{E}(z)=\tilde{\chi}(z) \circledast \tilde{E}(z).
\end{align}
It should be emphasized, that the off-diagonal components of $\hat{\tilde{M}}(z)$ describe scattering, i.e. the transfer of spectral amplitude between different wave vector components. 
The direct numerical integration of Equation~\eqref{eq:prop_fourierspace_eq} is possible, but numerically pathologically expensive. To motivate an essential next step of simplification we rewrite Equation~\eqref{eq:prop_fourierspace_eq} as
\begin{align}
    \underbrace{\left(\frac{\partial}{\partial z} - i\sqrt{k_z^2 +k_0^2\hat{\tilde{M}}(z)}\right)}_{\text{forward}}\underbrace{\left(\frac{\partial}{\partial z} + i\sqrt{k_z^2 +k_0^2\hat{\tilde{M}}(z)}\right)}_{\text{backward}}\tilde{E}(z)=0, 
\end{align}
where the differential equation is represented by a product of two operators associated with forward and backward propagating waves, respectively. 
This is how far we can get in describing Equation~\eqref{eq:prop_formally} in full generality. 

The next step towards a practical approximation results from the assumption that back-propagating waves can be neglected. In this case we can uniquely specify the axial wave vector as $k_z(k_x,k_y)=\sqrt{k_0^2 - k_x^2 - k_y^2}$ and obtain the tremendously simplified propagation equation
\begin{align}
    \left(\frac{\partial}{\partial z} - ik_z\sqrt{1 +\frac{k_0^2}{k_z^2}\hat{\tilde{M}}(z)}\right)\tilde{E}(z)=0.\label{eq:forward_propagation}
\end{align}
This equation is exact for the forward propagating waves and describes the spatial evolution of $\tilde E$ along the optical axis. For deriving practically useful schemes for its accurate  solution, it will turn out to be particularly convenient to introduce a contact transformation $\tilde {\bar E}=\frac{k_z}{k_0}\tilde E$, where $\tilde E$ is scaled by the obliquity factor $k_z/k_0$ known from e.g. Fraunhofer diffraction. As a result of this step, $\tilde{\bar E}(k_x,k_y;z)$ describes directly the amplitude of the plane waves associated with $(k_x,k_y)$, independent of the particular choice of the optical axis. The corresponding propagation equation follows from Equation~\eqref{eq:forward_propagation} as
\begin{align}
    \left(\frac{\partial}{\partial z} - ik_z\frac{k_z}{k_0}\sqrt{1 +\frac{k_0^2}{k_z^2}\hat{\tilde{M}}(z)}\,\frac{k_0}{k_z}\right)\tilde{\bar E}(z)=0.
    \label{eq:forward_propagation_w_obliquity}
\end{align}
In the spirit of the WKB-ansatz we now translate the differential equation into a propagator form using a phase integral representation
according to
\begin{align}
    \tilde {\bar E}(z)=\underbrace{e^{i{\int_{z_0}^z \hat {\tilde T}(z')\, dz'}}}_{\hat{\tilde{P}}(z)} \tilde{\bar E}(z_0).
    \label{eq:WKB_ansatz}
\end{align}
Here, the propagator $\hat{\tilde{P}} (z)$ maps the initial field $\tilde{\bar E}(z_0)$, sampled at axial position $z_0$, to the field $\tilde{\bar E}(z)$ at target position $z$. Therein, the operator $\hat{\tilde{T}}(k_x, k_y; z')$ can be identified by inserting  Equation~\eqref{eq:WKB_ansatz} in Equation~\eqref{eq:forward_propagation_w_obliquity} as
\begin{align}
\hat {\tilde \wavtrans}(z)=k_z\frac{k_z}{k_0}\sqrt{1+\hat{\tilde{X}}(z)}\frac{k_0}{k_z}, \quad \text{with} \quad \hat{\tilde{X}}(z) = \frac{k_0^2}{k_z^2}\hat{\tilde{M}}(z)=\frac{k_0^2}{k_z^2}\mathcal{ \hat F} \chi(z)\mathcal{\hat 
 F}^{-1}.
\label{eq:wkb_helmholtz_propagator}
\end{align}
The diagonal of the operator $\hat {\tilde{T}}(z)$ has the role of an effective axial wave number, while non-diagonal terms describe the transfer between wave vector components. For the trivial case, in the absence of a medium ($\hat{\tilde{M}}(z)=\hat{0}$), we obtain the vacuum solution $\hat{\tilde{\wavtrans}}_\text{vac}=k_z(k_x, k_y)$, which is equivalent to the well known angular-spectrum-method~\cite{Goodman2005}. In the general case, i.e. for non-zero $\hat{\tilde{M}}(z)$, the additional effect of the auxiliary operator $\hat{\tilde{X}}(z)$ in Equation~\eqref{eq:wkb_helmholtz_propagator} can be described by a Taylor expansion of the square root as
\begin{align}
\hat {\tilde \wavtrans}(z)&=k_z+k_z\frac{k_z}{k_0}\sum_{m=1}^\infty \binom{1/2}{m} \hat{\tilde{X}}^m(z) \,\frac{k_0}{k_z}=\hat {\tilde \wavtrans}_{\text{vac}}+\hat {\tilde \wavtrans}_{\text{mat}}(z),
\end{align}
which reflects a representation that can, in principle, be evaluated explicitly. 
Here, we identify the term $\hat {\tilde \wavtrans}_\text{mat}(z)$ as the material correction  with
\begin{align}
\hat {\tilde \wavtrans}_\text{mat}(z)&=k_z\frac{k_z}{k_0}\sum_{m=1}^\infty \binom{1/2}{m} \biggl[\underbrace{\frac{k_0^2}{k_z^2}\mathcal{ \hat F} \chi(z)\mathcal{\hat  F}^{-1}}_{\hat{\tilde{X}}(z)}\biggr]^m \frac{k_0}{k_z}. \label{eq:formal_tmat}
\end{align}
A  brute-force-evaluation of the $m$-th power of $\hat{\tilde{X}}(z)$, in general, would require $m$ pairs of Fourier transforms because $k_z(k_x,k_y)$ is defined in $k$-space and $\chi(x,y;z)$ needs to appear in real-space. For an efficient practical treatment the material correction can be evaluated within the small-angle approximation $(k_z\approx k_0)$, which we henceforward indicate by a superscript ''sa''. As a result, the dependence on $k_z$ is removed and the Fourier transforms can be pulled out of the term in square brackets, since adjacent regular and inverse Fourier transforms cancel in higher orders of $m$. We thus obtain
\begin{align}
    \hat{\tilde{\wavtrans}}_\text{mat}^\text{sa}(z) &= k_0\mathcal{ \hat F}\sum_{m=1}^\infty \binom{1/2}{m} \chi^m(z)\mathcal{\hat  F}^{-1}=k_0 \mathcal{ \hat F}\big(\underbrace{\sqrt{1+\chi(z)}}_{n(z)}-1\big)\mathcal{\hat  F}^{-1}  \label{eq:Tmat_approx}
\end{align}
and emphasize that this approximation even matches the linear term ($m=1$) in the expansion in Equation~\eqref{eq:formal_tmat}  when applied to fields with $k_z=k_0$. Hence, single scattering of the primary beam (with wave vector along the optical axis) is described exactly. 

At this point we are finally able to evaluate the field propagation according to Equation~\eqref{eq:WKB_ansatz} using a step-wise integration over thin slices $\Delta z$
\begin{align}
    \tilde{\bar E}(z+\Delta z)=e^{i (\hat {\tilde \wavtrans}_{\text{vac}}+\hat {\tilde \wavtrans}_{\text{mat}}^{\rm sa}(z)) \Delta z}\tilde{\bar E}(z),
\end{align}
which is justified if the slice thickness $\Delta z$ is sufficiently small such that the $z$-dependence of $n(z)$ and thus also of $\hat{\tilde{T}}(z)$ within this slice can be neglected. Furthermore, we can factorize the exponential according to the Baker-Campbell-Haussdorff formula as
\begin{align}
    \tilde{\bar E}(z+\Delta z)=e^{i\hat{\tilde{T}}_{\rm vac} \Delta z}e^{i\hat {\tilde \wavtrans}_{\text{mat}}^{\rm sa}(z) \Delta z}\underbrace{e^{[\hat{\tilde{T}}_{\rm vac}, \hat{\tilde{T}}_{\rm mat}^{\rm sa}(z)]\Delta z^2/2}}_{\hat{\tilde{W}}(z)} \tilde{\bar E}(z),
\end{align}
where the operator $\hat{\tilde{W}}(z)$ describes the effect of multiple scattering within a single slice. The Taylor expansion
\begin{align}
    \hat{\tilde{W}}(z) = 1+[\hat{\tilde{T}}_{\rm vac}, \hat{\tilde{T}}_{\rm mat}^{\rm sa}(z)] \Delta z^2/2 + O(\Delta z ^4)
\end{align}
shows that we can assume the contribution from multiple scattering within the slice to vanish ($\hat{\tilde{W}} = \hat 1$) if $\Delta z$ is chosen sufficiently small. Note that if $\hat{\tilde{T}}_{\rm vac}$ and $\hat{\tilde{T}}_{\rm mat}^{\rm sa}(z)$ commute, as realized e.g. in homogeneous media, this factorization is even exact for arbitrarily large $\Delta z$.

Using sufficiently thin slices, the factorization allows us to write the propagator as
\begin{align}
    \hat{\tilde{P}}(z) = \hat{\tilde{P}}_{\rm vac} \hat{\tilde{P}}_{\rm mat}^{\rm sa}(z), \quad \text{with} \quad \hat{\tilde{P}}_{\rm vac}=e^{i\hat{\tilde{T}}_{\rm vac} \Delta z}=e^{i k_z\Delta z} \quad \text{and} \quad \hat{\tilde{P}}_{\rm mat}^{\rm sa}(z)=e^{i\hat {\tilde \wavtrans}_{\text{mat}}^{\rm sa}(z) \Delta z},\label{eq:pMSFT_vacpropagators_definition}
\end{align}
where $\hat{\tilde{P}}_{\rm vac}$ and $\hat{\tilde{P}}_{\rm mat}^{\rm sa}(z)$ describe the vacuum propagation and material correction, respectively. The latter can be rewritten  conveniently by pulling the Fourier transforms contained in $\hat {\tilde \wavtrans}_{\text{mat}}^{\rm sa}(z)$ out of the exponent. The justification for that is most easily seen when Taylor-expanding the exponential according to 
\begin{align}
    \hat{\tilde{P}}_{\rm mat}^{\rm sa}(z)=e^{i\hat {\tilde \wavtrans}_{\text{mat}}^{\rm sa}(z) \Delta z}&=\sum_{m=0}^\infty \frac{1}{m!}\left( \mathcal{ \hat F} i k_0 \big(n(z)-1\big) \Delta z \mathcal{\hat  F}^{-1}\right)^m=\mathcal{ \hat F} \underbrace{e^{i k_0 \left(n(z)-1\right)\Delta z}}_{\hat{P}_{\rm  mat}^{\rm sa}(z)}\mathcal{\hat  F}^{-1},
\end{align}
where we have exploited again the cancellation effect of adjacent Fourier transforms for higher orders and have expressed the material correction directly in real-space via $\hat{P}_{\text{mat}}^{\text{sa}}(z)$. The final form of the propagation over one slice yields a split-step representation of vacuum propagation and material correction with 
\begin{align}
    \underbrace{\tilde{\bar E}(z+\Delta z)}_{\frac{k_z}{k_0}\tilde E(z+\Delta z)} &= \underbrace{e^{ik_z\Delta z}}_{\hat{\tilde{P}}_{\rm vac}}\, \mathcal{ \hat F} \underbrace{e^{i k_0 \left(n(z)-1\right)\Delta z}}_{\hat{P}_{\rm  mat}^{\rm sa}(z)}\mathcal{\hat  F}^{-1}\underbrace{\tilde{\bar E}(z)}_{\frac{k_z}{k_0} \tilde E(z)} = \hat{\tilde{P}}_{\rm vac}\hat{\tilde{P}}_{\rm mat}^{\rm sa}(z)\tilde{\bar E}(z). \label{eq:split_step}
\end{align}
Note that this split-step formulation utilizing exponential functions yields unconditional  numerical stability. A single propagation step according to Equation~\eqref{eq:split_step} requires only two transverse Fourier-transforms and trivial multiplications with complex amplitude maps in real and k-space, respectively, and can therefore be implemented in a straight-forward way numerically.

Employing the method to the field propagation through an extended sample requires a spatial discretization of the object into $N$ slices and the successive application of the above propagator for each slice $s$ according to
\begin{equation}
    \tilde{\bar E}(N\Delta z) = \left[\prod_{s=1}^N \hat{\tilde{P}}_{\rm vac}\hat{\tilde{P}}_{\rm mat}^{\rm sa}(s\Delta z)\right]\tilde{\bar E}(0). 
    \label{eq:total_exit_field}
\end{equation}
We term this approach the propagation multi-slice Fourier-transform (pMSFT) method. Note that the ordering in the chain of operators resulting from the product is important and must reflect forward propagation (in $\vec e_z$ direction).
Equation~\eqref{eq:total_exit_field} yields the total exit field behind the target that contains a superposition of the unperturbed incident field and the scattered field. In cases, where only the scattered field $\tilde{\bar E}_{S}$ is of interest, the hypothetical vacuum-propagated incident field must be subtracted from the total field via
\begin{align}
    \tilde{\bar E}_{S} &= \left[\prod_{s=1}^N \hat{\tilde{P}}_{\rm vac}\hat{\tilde{P}}_{\rm mat}^{\rm sa}(s\Delta z)\right]\tilde{\bar E}(0) -\left[\prod_{s=1}^N \hat{\tilde{P}}_{\rm vac}\right]\tilde{\bar E}(0).\label{eq:scattered_exit_field}
\end{align}
This represents the pMSFT prediction of the scattered field  $\tilde{\bar E}_{S}(k_x,k_y)$ and is the main result of this section. The final step is the propagation to the detector in the far-field and is discussed in Section \ref{sec:ff_detector}.

\section{Connection to established methods\label{sec:pMSFTconnection}}
The pMSFT algorithm represents a split-step-based solution to the scattering problem with minimal approximations. In the following, a chain of additional approximations will be discussed that allows the step-by-step derivation of all the other established methods mentioned in the introduction, i.e. Hare's paraxial split-step\\method, MSFT, Born and SAXS. This procedure provides a complete physical picture of the nature of the approximations underlying each individual method as summarized in \textbf{Table~\ref{tab:eq_table}}. 

We begin with the connection to Hare's paraxial split-step approach~\cite{Hare1994}, which follows from pMSFT by applying the paraxial approximation to the vacuum propagator in Equation~\eqref{eq:pMSFT_vacpropagators_definition}. To this end we apply a  linear Taylor expansion to the square root in the axial wave number according to 
\begin{align}
 k_z(k_x,k_y) = \sqrt{k_0^2 - k_x^2 - k_y^2} \approx k_0 - \frac{k_x^2+k_y^2}{2k_0},
\end{align}
which is valid in the paraxial limit. Inserting the resulting paraxial vacuum propagator 
\begin{align}
 \hat{\tilde{P}}_{\rm vac}^{\rm parax} = e^{i\Delta z k_0}e^{-i\Delta z \frac{k_x^2+k_y^2}{2k_0}} \label{eq:paraxial_vacuum_propagator}
\end{align}
in Equation~\eqref{eq:scattered_exit_field} yields the scattered exit field
\begin{align}
    \tilde{\bar E}_{S}^{\rm parax}=\left(\left[\prod_{s=1}^N \hat{\tilde{P}}_{\rm vac}^{\rm parax}\hat{\tilde{P}}_{\rm mat}^{\rm sa}(s\Delta z)\right] - \left[\prod_{s=1}^N\hat{\tilde{P}}_{\rm vac}^{\rm parax}\right]\right)\tilde{\bar E}(0). \label{eq:paraxial_scattered_exitfield}
\end{align}
This form is equivalent to Hare's approach~\cite{Hare1994}, which was motivated ad-hoc without a rigorous derivation.
Note that Equation~\eqref{eq:paraxial_scattered_exitfield} augments Hare's result by the missing prefactor needed for quantitative predictions of scattered fields.
Surprisingly, despite its merits, Hare's method has, to the best of our knowledge, not been used systematically in scattering simulations so far. However, when compared directly to pMSFT, the paraxial split-step method in Equation~\eqref{eq:paraxial_scattered_exitfield} is not competitive as it delivers a less accurate result (cf. Section~\ref{sec:benchmark}) at the same numerical cost and complexity.

To establish a physically transparent connection to the remaining other methods we switch from the so-far-used propagation representation (cf. Equation~\eqref{eq:scattered_exit_field}) to a multi-slice representation. This will allow us to write the scattered field in the exit plane as a superposition of contributions from individual slices. To this end we rewrite the first term of Equation~\eqref{eq:scattered_exit_field} that contains an alternating product of non-commuting operators. Using generic operators $\hat{a}$ and $\hat{b}_i$, such a product can be rewritten as
\begin{align}
    \prod_{i=1}^N \hat{a}\hat{b}_i&= \left(\hat{a} \left(\hat{b}_N - 1\right) + \hat{a}\right)\prod_{i=1}^{N-1} \hat{a}\hat{b}_i \label{eq:generic_operators_idea}\\
    &= \hat{a} \left(\hat{b}_N - 1\right)\prod_{i=1}^{N-1} \hat{a}\hat{b}_i + \hat{a} \prod_{i=1}^{N-1} \hat{a}\hat{b}_i,\label{eq:generic_operators_idea_exec}
\end{align}
where the product of operators on the right hand side of Equation~\eqref{eq:generic_operators_idea} has the same structure as the product on the left hand side, up to a simple index shift. This feature enables the recursive insertion of Equation~\eqref{eq:generic_operators_idea_exec} into the last term on the right hand side of Equation~\eqref{eq:generic_operators_idea_exec} . The chain is terminated in the last step by $\hat a\hat b_1=\hat a (\hat b_1-1) +\hat a$ and yields
\begin{align}
    \prod_{i=1}^N \hat{a}\hat{b}_i=\sum_{i=1}^N\left(\left[\prod_{k=i}^N\hat{a}\right]\left(\hat{b}_i-1\right)\left[\prod_{l=1}^{i-1} \hat{a}\hat{b}_l\right]\right) + \left[\prod_{i=1}^N\hat{a}\right]. \label{eq:generic_operators_done}
\end{align}
Associating the operators $\hat a$ and $\hat b_i$ with $\hat{\tilde{P}}_{\rm vac}$ and $\hat{\tilde{P}}_{\rm mat}^{\rm sa}(i\Delta z)$ and using Equation~\eqref{eq:generic_operators_done} to rewrite  the first term of Equation~\eqref{eq:scattered_exit_field} leads to the fully equivalent multi-slice representation
\begin{align}
    \tilde{\bar E}_{S}=\sum_{s=1}^N\Biggl(\underbrace{\left[\prod_{k=s}^N\hat{\tilde{P}}_{\rm vac}\right]}_{\text{path to exit}}\underbrace{\left(\hat{\tilde{P}}_{\rm mat}^{\rm sa}(s\Delta z)-1\right)}_{\text{scatt. strength of slice}}\underbrace{\left[\prod_{l=1}^{s-1} \hat{\tilde{P}}_{\rm vac}\hat{\tilde{P}}_{\rm mat}^{\rm sa}(l\Delta z)\right]}_{\text{path to slice}}\tilde{\bar E}(0)\Biggr). \label{eq:multislice_rep}
\end{align}
Each term of this sum can further be separated into factors that are associated with specific stages of the scattering process, namely, with the propagation to the slice, the scattering strength of the slice, and the propagation to the exit plane, respectively (as indicated). This allows us to introduce approximations to individual stages that establish the links to the remaining methods (MSFT, Born, and SAXS). The key steps of approximation and the associated propagation equations are given in Table~\ref{tab:eq_table}.

First, we apply the small-angle approximation to the vacuum propagator in the "path to slice" term, which then becomes a constant $\hat{\tilde{P}}_{\rm vac}^{\rm sa}=e^{i k_0\Delta z}$. Only in the case that the incident field is a plane wave propagating along the optical axis, is its vacuum propagation still exact within this approximation.

As the propagators $\hat{\tilde{P}}_{\rm vac}^{\rm sa}$ and $\hat{\tilde{P}}_{\rm mat}^{\rm sa}(z)$ now commute, we can separate them in the path towards the slice with index $s$ according to
\begin{align}
    \left[\prod_{l=1}^{s-1}\hat{\tilde{P}}_{\rm vac}\hat{\tilde{P}}_{\rm mat}^{\rm sa}(l\Delta z)\right]\tilde{\bar E}(0)&\approx \left[\prod_{l=1}^{s-1}\hat{\tilde{P}}_{\rm vac}^{\rm sa}\right] \left[\prod_{l=1}^{s-1}\hat{\tilde{P}}_{\rm mat}^{\rm sa}(l\Delta z)\right]\tilde{\bar E}(0)\\
    &=\hat{\mathcal{F}}\left(e^{i \Delta z k_0 \sum_{l=1}^{s-1}\left(n(l \Delta z)\right)}\right)\hat{\mathcal{F}}^{-1}\tilde{\bar E}(0) =\hat{\mathcal{F}}\left(e^{i \Delta z k_0 \sum_{l=1}^{s-1}\left(n(l \Delta z)\right)}\right)\hat E , \label{eq:msft_pathtoslice}
\end{align}
where we have exploited that adjacent Fourier transforms cancel and that the real-space representation of a plane wave in the initial slice is a constant amplitude $\hat E$. Here, the sum over the slice-dependent refractive index in the exponential function reflects a material projection and resembles the well-known Beer-Lambert law.

This treatment effectively describes absorption and wave front deformation, but completely neglects refraction and diffraction on the path towards a slice.
The remaining stages of the scattering process, i.e. the diffraction within a slice and the propagation to the exit plane are treated analogously to pMSFT.
Describing the path-to-slice term in Equation~\eqref{eq:multislice_rep} with the result of Equation~\eqref{eq:msft_pathtoslice} yields a description that is conceptually equivalent to multi-slice Fourier-transform (MSFT) methods that have been employed systematically to soft X-ray single particle diffraction previously~\cite{Barke2015, Langbehn2018, Colombo2022}. Moreover, our rigorous derivation of the MSFT propagation equation, see Table~\ref{tab:eq_table}, even allows quantitative field predictions and contains the most general form of the scattering strength that is valid also for large departures of the refractive index from unity. The latter  feature is particularly important for applications to heterogeneous objects containing regions with different optical properties. Note that a Taylor expansion of the scattering strength used in Ref.~\cite{Colombo2022} around $n=1$ matches our results only up to the linear term, limiting the correct description of relative material contrast to small departures of the refractive index.

Next, we consider the limiting case of very weak scattering ($n\approx1$). To this end we keep the full vacuum propagator but neglect the material correction in the path-to-slice term, which becomes 
\begin{align}
    \left[\prod_{l=1}^{s-1} \hat{\tilde{P}}_{\rm vac}\hat{\tilde{P}}_{\rm mat}^{\rm sa}(l\Delta z)\right]\tilde{\bar E}(0)\approx \left[\prod_{l=1}^{s-1} \hat{\tilde{P}}_{\rm vac}\right]\tilde{\bar E}(0)= e^{i (s-1)\Delta z k_z }\tilde{\bar E}(0).
\end{align}
After inserting this into Equation~\eqref{eq:multislice_rep} we obtain the  scattered fields in first Born approximation~\cite{Born1926}, see Table~\ref{tab:eq_table}. Therein, scattering in each slice is evaluated using the  unperturbed incident field, fully neglecting the material effects on the propagation before and after scattering.
\begin{table}[ht]
\caption{\label{tab:eq_table} Summary of propagation equations underlying the different simulation methods together with a short description of the corresponding approximations with respect to the propagation multi-slice Fourier transform method. }
\begin{center}
\begin{tabular}{c|lr}
    Method&discretized exit field equation \hfill &\hspace{-6em}approximation with respect to pMSFT \\
    \hline\\[-8pt]
    
    \multirow{2}{*}{pMSFT}&\multirow{2}{*}{\(\displaystyle\tilde{\bar E}_{S}^{\rm pMSFT} = \left[\prod_{s=1}^N e^{i\Delta z k_z} \mathcal{\hat F}e^{i\Delta zk_0\left(n(s\Delta z)-1\right)} \mathcal{\hat F}^{-1}\right]\tilde{\bar E}(0) -e^{iN\Delta z k_z}\tilde{\bar E}(0)\)}&\multirow{2}{*}{-}\\
    &&\\[8pt]
    
    Hare's&\multirow{2}{*}{\(\displaystyle \tilde{\bar E}_{S}^{\rm parax} = e^{iN\Delta z k_0}\left[\prod_{s=1}^N e^{-i\Delta z \frac{k_x^2+k_y^2}{2k_0}}\hat{\mathcal{F}}e^{i\Delta zk_0\left(n(s\Delta z)-1\right)}\hat{\mathcal{F}}^{-1}\right]\tilde{\bar E}(0) -e^{iN\Delta z k_0}e^{-iN\Delta z \frac{k_x^2+k_y^2}{2k_0}}\tilde{\bar E}(0)\)}&\multirow{2}{*}{all $\displaystyle k_z \approx k_0 - \frac{k_x^2+k_y^2}{2k_0}$}\\
    paraxial&&\\[8pt]

    \multirow{2}{*}{MSFT}&\multirow{2}{*}{\(\displaystyle \tilde{\bar E}_{S}^{\rm MSFT} =\sum_{s=1}^N\left(e^{i(N-s)\Delta z k_z}\hat{\mathcal{F}}\left(e^{i\Delta zk_0\left(n(s\Delta z)-1\right)}-1\right)e^{i \Delta z k_0 \sum_{l=1}^{s-1}\left(n(l \Delta z)\right)}\right)\hat E\)}&\(\displaystyle k_z \approx k_0\) along inward path\\
    &&\hspace{-10em}incident plane wave with amplitude $\hat E$\\[8pt]

    \multirow{2}{*}{Born}&\multirow{2}{*}{\(\displaystyle \tilde{\bar E}_{S}^{\rm Born} =\sum_{s=1}^N\left(e^{i(N-s)\Delta z k_z}\hat{\mathcal{F}}\left(e^{i\Delta zk_0\left(n(s\Delta z)-1\right)}-1\right)\hat{\mathcal{F}}^{-1}e^{i(s-1)\Delta z k_z}\right)\tilde{\bar E}(0)\) }&\multirow{2}{*}{\(\displaystyle (n-1) \approx 0\) along inward path}\\
    &&\\[8pt]

    \multirow{2}{*}{SAXS}&\multirow{2}{*}{\(\displaystyle \tilde{\bar E}_{S}^{\rm SAXS} = e^{iN\Delta z k_0}\hat{\mathcal{F}}\left(\sum_{s=1}^N\left(e^{i \Delta z k_0\left(n(s \Delta z)-1\right)}-1\right)\right)\hat E\)}&\hspace{-99em}all \(\displaystyle k_z \approx k_0\) and \(\displaystyle (n-1) \approx 0\) along inward path\\
    &&\hspace{-10em}incident plane wave with amplitude $\hat E$\\
\end{tabular}
\end{center}
\end{table}
\ifdefined\printImages
\begin{figure}[!ht]
    \centering
    \includegraphics[width=1\hsize]{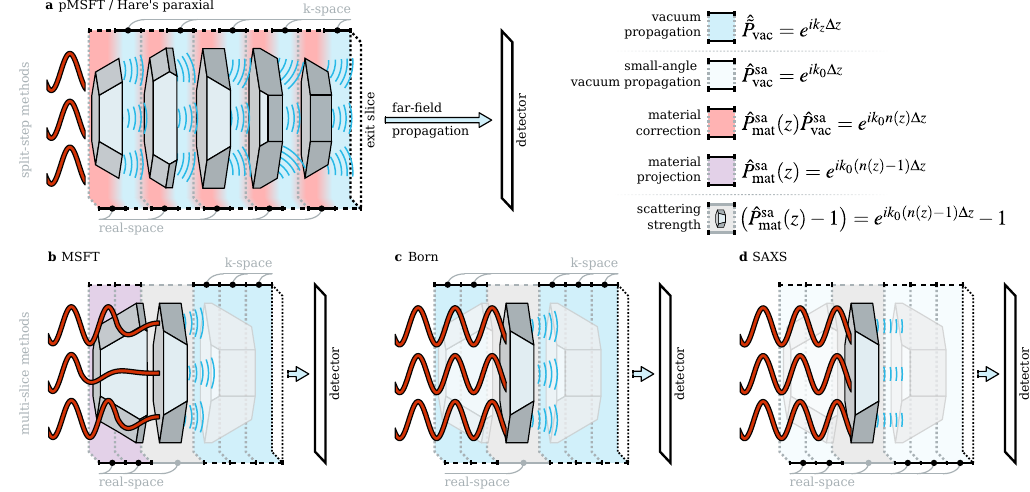}
    \caption{Schematic overview of different diffraction methods. (a) Propagation multislice Fourier transform (pMSFT) method with a series of alternating sub-steps representing material correction in real space (red) and vacuum propagation in k-space (blue) for each slice.  (b) Conventional MSFT method with Beer-Lambert-type material projection (lilac) in real space on the path to the slice. After application of the scattering strength (grey) in the slice the resulting field is vacuum-propagated (material neglected) to the exit plane in k-space. (c) Born method with vacuum propagation to the slice in k-space, application of the scattering strength in the slice in real space followed by vacuum propagation in k-space to the exit plane. (d) SAXS approach with pure summation of the scattering strength over all slices resulting in a single projected scattering strength evaluated in the exit plane. For all cases the resulting exit field can be propagated to the detector using a far field transformation.}    \label{fig:msft_schemes}
\end{figure}
\fi
Employing the small-angle approximation to the Born result, i.e. $k_z\approx k_0$ and restricting the incident field to a plane wave with amplitude $\hat E$, we obtain
\begin{align}
\tilde{\bar E}_{S}^{\rm SAXS}&\approx\sum_{s=1}^N\left(\left[\prod_{k=s}^N\hat{\tilde{P}}_{\rm vac}^{\rm sa}\right]\left(\hat{\tilde{P}}_{\rm mat}^{\rm sa}(s\Delta z)-1\right)\left[\prod_{l=1}^{s-1} \hat{\tilde{P}}_{\rm vac}^{\rm sa}\right]\tilde{\bar E}(0)\right)\\
&=e^{iN\Delta z k_0}\mathcal{\hat{F}}\left(\sum_{s=1}^N\left(e^{i \Delta z k_0\left(n(s \Delta z)-1\right)}-1\right)\right)\hat E,
\end{align}
which corresponds to the small-angle X-ray scattering (SAXS) approximation~\cite{Guinier1979}, the standard method for hard X-ray diffraction. 
In this limit, material only plays a role in the determination of the scattering strength, while propagation is described using only the small-angle vacuum propagator and can therefore be represented by a single global phase factor in front of the propagation equation. As a result, the exit field is determined by the sum of scattering strengths, i.e. their projection. To illustrate the nature of the approximations underlying the specific methods more intuitively,  the physics represented by the respective propagation equations is sketched schematically in \textbf{Figure~\ref{fig:msft_schemes}}.

Common to the numerical implementation of the above methods is the sampling of both the objects shape and its optical properties via the refractive index map on a three dimensional grid representing the $N$ two-dimensional slices. In order to take advantage of Fast-Fourier-Transform techniques~\cite{Press2007} it is particularly convenient to use an equidistant grid spacing. The numerical implementation of pMSFT requires the execution of an ordered sequence of alternating vacuum propagation and material correction steps with interleaved Fourier transforms. As a result, the involved $2N$ Fourier transforms are not independent from each other, impeding an acceleration by parallel processing of individual slices. 
For MSFT, the material projection for an incident plane wave, the evaluation of the scattering strength, and the following vacuum propagation can be described for each slice individually, allowing parallel processing of the $N$ independent Fourier transforms. For the Born approximation, both the vacuum propagation to the slice and from the slice to the exit plane have to be performed in k-space while the scattering strength has to be evaluated in real-space. As a result, Born requires in general $2N$ independent Fourier transforms that are suited for parallel computing and can even be reduced to $N$ transforms in the case of an incident plane wave. Finally, the SAXS method consists of a mere projection of the scattering strength followed by a single Fourier transformation and is thus extremely fast.

In conclusion, pMSFT, MSFT, and Born have the same numerical complexity, while MSFT and Born may benefit from parallelization of an individual scattering simulation. However, this seeming drawback of pMSFT is typically irrelevant in performance-limited practical applications, such as the forward fitting of experimental images~\cite{Barke2015, Rupp2017, Langbehn2018, Colombo2023}. Commonly employed optimization techniques (e.g. Simplex~\cite{Nelder1965} and genetic algorithms~\cite{Mitchell1998}) are associated with the simulation of a swarm of independent geometry candidates for each iteration, which can be parallelized perfectly.

\section{Prediction of the scattering cross section and the far-field measured by a detector} \label{sec:ff_detector}
We now return to the task of propagating the resulting scattered exit field $\tilde{\bar E}_S(k_x, k_y)$ to a detector. Note that this process is identical for all methods discussed above and can be performed analogously for the total exit field.

The seemingly obvious approach of using the exact vacuum propagation via $\hat{\tilde{P}}_\text{vac}$ from Equation~\eqref{eq:pMSFT_vacpropagators_definition} is highly inconvenient as this would require the size of the transverse grid to match the detector dimension. The two commonly used approximate methods for the propagation to a distant detector are the Fresnel and the Fraunhofer transformations, where the latter represents the stronger approximation with higher requirements on the detector distance~\cite{Altenkirch2024}. 
For the envisioned application to XUV and X-ray scattering, the Fraunhofer transformation is typically justified. Therein, the resulting complex electric far-field at an arbitrary point $\vec r$ follows from the scattered exit field $\tilde{\bar E}_S(k_x, k_y)$ as
\begin{align}\label{eq:fraunhofer_scatteredfield}
    E(\vec r)&=\tilde{\bar E}_S(k_x, k_y) \, \frac{k_0}{i} \frac{e^{i\vec k \cdot \vec r}}{|\vec r|}, \quad \mbox{with} \quad 
    \vec{r}=\begin{pmatrix} x \\ y \\ z\end{pmatrix}  \quad  \mbox{and} \quad  \vec{k}=\begin{pmatrix}k_x \\ k_y \\ k_z\end{pmatrix}=k_0\frac{\vec{r}}{|\vec r|},
\end{align}
where the center of the corresponding coordinate system is considered at the intersection of the optical axis with the exit field slice. We like to emphasize that the common form of the Fraunhofer transformation contains a product of $\tilde{E}$ and the obliquity factor $k_z/k_0$, or $z/r$, which corresponds to $\tilde{\bar{E}}$, illustrating the central role of $\tilde{\bar E}$ as the physically relevant field amplitude. 
Note that the description presented so far is based on the scalar wave equation and thus does not fully account for polarization effects. This shortcoming is most pronounced for scattering to large angles and can be effectively corrected for in the case of an incident plane wave in single scattering approximation (see supporting information). The resulting angular dependent correction mask for the scattered electric field in the case of incident plane wave polarized along the y-axis reads
\begin{equation}
\Gamma(k_x, k_y)=\sqrt{1-\frac{k_y^2}{k_0^2}}.
\end{equation}
In order to employ this correction, $\tilde{\bar E}_S(k_x, k_y)$ has to be replaced by $\Gamma(k_x,k_y)\tilde{\bar E}_S(k_x, k_y)$ in Equation~\eqref{eq:fraunhofer_scatteredfield}. 

The remaining task is to translate the far-field  to the practically measurable power density by taking into account the geometry and the local tilt of the detector. To this end we depart from the total scattered flux resulting from the integration of the Poynting vector $\vec S(\vec r) = \frac{\epsilon_0 c_0}{2} |E(\vec r)|^2\vec e_r$ over a closed surface $(V)$ with oriented surface elements $d\vec{A}(\vec r)$ with
\begin{align}
    \Phi_{tot} = \int_{(V)} \vec S (\vec r)\cdot d\vec{A}(\vec r).  \label{eq:total_flux}
\end{align}
Relating the flux to the incident intensity $I_0 = \frac{\epsilon_0 c_0}{2} \left|E_0\right|^2$ yields the total scattering cross section
\begin{align}
    \sigma_{\rm tot} = \frac{\Phi_{tot}}{I_0} = \frac{1}{I_0} \int_{(V)} \vec S (\vec r)\cdot d\vec{A}(\vec r).
    \label{eq:total_csec}
\end{align}
A particularly simple structure of the integral follows for the case of a spherical detector with radius $R$ and surface element $d \vec A_s=R^2\, d\Omega \,\vec e_r$ via
\begin{align}
    \sigma_{tot} = \int \frac{d\sigma}{d\Omega} d\Omega, \qquad \mbox{with} \qquad 
    \frac{d\sigma}{d\Omega}=\Gamma^2\frac{|\tilde{\bar E}_S|^2}{|E_0|^2} \, k_0^2,
    \label{eq:sigma_tot_sigma_diff}
\end{align}
where we introduced the differential scattering cross section $\frac{d\sigma}{d\Omega}$. For comparison with experiments a given element of solid angle $d\Omega$ can be associated with surface elements $dA_s$ on a spherical detector as above or alternatively with $dA_f=dx\,dy$ on a flat detector at distance $R$ or with $dA_k=dk_x\,dk_y$ in k-space  via
\begin{align}
    d\Omega = \frac{1}{R^2}\, dA_s = \frac{\cos^3\vartheta}{R^2}\, dA_f = \frac{1}{k_0^2\cos\vartheta}\, dA_k, \quad\text{with} \quad\cos \vartheta=\frac{k_z}{k_0}.
\end{align}
The specific pre-factors represent the corresponding Jacobi determinants resulting from the associated coordinate transformation. 
For example, the differential cross section with surface elements in k-space follows as
\begin{align}
    \frac{d \sigma}{d A_k} =\Gamma^2\frac{\tilde{\bar E}_S^2(k_x, k_y)}{\left|E_0\right|^2} \, \frac{1}{\cos \vartheta},
\end{align}
where both the scattering cross section and the scattered exit field $\tilde{\bar E}_S(k_x, k_y)$ are represented in a common Cartesian coordinate system $(k_x, k_y)$. The latter fact and the absence of any parameters of the setup (e.g. detector distance or tilt) makes this representation universal and particularly convenient for comparison and graphical visualization. The translation into the associated power density measured in a practical experiment is straightforward  and, for the example of a flat detector, leads to
\begin{align}
    \frac{d\Phi}{dA_f}=I_0\frac{d\sigma}{dA_f} = I_0 \frac{d \sigma}{d A_k} \frac{d A_k}{d A_f} = I_0 \frac{d \sigma}{d A_k} \frac{k_0^2\cos^4 \vartheta}{R^2} 
\end{align}
and allows the re-sampling of experiment and theory on a common grid  for quantitative comparison. The treatment of other detector shapes is analogous. 
Note that the polarization correction is used for all results shown in this manuscript.  It can be excluded by simply setting $\Gamma$ to unity.
\FloatBarrier
\section{Method benchmark}\label{sec:benchmark}
The qualitative comparison of diffraction images predicted by the different methods as displayed in Figure~\ref{fig:fdtd_vs_pmsft} for a truncated silver octahedron has shown the exquisite match of the pMSFT prediction with the FDTD result and illustrates the systematic loss of feature reproduction capabilities with increasing degree of approximation. In the following we present a comprehensive quantitative quality benchmark. To this end we consider optical properties most relevant for diffraction experiments in the spectral range from the UV to hard X-rays as illustrated in \textbf{Figure~\ref{fig:benchmark_scenario}}. Figure~\ref{fig:benchmark_scenario}~\hyperref[fig:benchmark_scenario]{(a)} and~\hyperref[fig:benchmark_scenario]{(b)} display the spectral evolution of the complex refractive index for typical sample materials. The representation of this refractive index data  in the complex plane in Figure~\ref{fig:benchmark_scenario}~\hyperref[fig:benchmark_scenario]{(c)} (colored lines) together with the optical properties of all bulk materials listed in the  CXRO-database by Henke \textit{et. al}~\cite{Henke1993} (gray lines) defines the relevant refractive index region considered in our comparison. Note that the largest departures of the refractive index from unity occur in the XUV range while a convergence towards unity takes place in the limit of hard X-rays.

\ifdefined\printImages
\begin{figure*}[!ht]
    \fontsize{8}{10}\selectfont
    \centering
    \includegraphics[width=1\hsize]{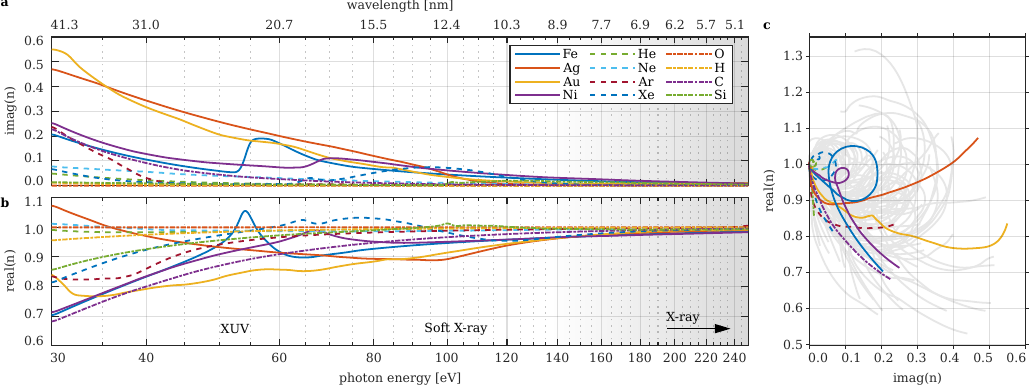}
    \caption{Frequency dependent refractive index for various materials from the extreme ultraviolet (XUV) to the X-ray regime. The evolution of real and imaginary parts of the refractive index for selected materials (as indicated) are shown in panels (a) and (b), respectively. The corresponding evolutions in the complex plane are displayed in (c) together with additional data for all materials listed in the CXRO database~\cite{Henke1993} (gray traces).}
    \label{fig:benchmark_scenario}
\end{figure*}
\fi
\noindent
A qualified assessment of the individual model predictions can be achieved by comparison of the respective far-field scattering images with an exact reference. The scattering of a plane wave with wavelength $\lambda$ by a homogeneous sphere with diameter $D$ is a convenient test scenario as an exact analytical reference is available via Mie's solution~\cite{Mie1908} in the full considered refractive index region. A suitable observable for our comparison is the differential scattered fraction, which relates the differential scattering cross section to the geometrical cross section of the sample $\sigma_{geo}$ (for a sphere $\sigma_{geo}=\pi D^2/4$) according to
\begin{align}
    \Lambda(\vartheta, \varphi) = \frac{1}{\sigma_\mathrm{geo}} \frac{d\sigma}{d\Omega}(\vartheta, \varphi).
\end{align}
The central advantage of the scattered fraction is the fact that for a given object shape it only depends on the object size to wavelength ratio (in our case $D/\lambda$) and the refractive index $n$ . This universal character applies to the Mie solution as well as to the results of all of the above discussed grid-based methods when analyzed using Equation~\eqref{eq:sigma_tot_sigma_diff}. In particular, the  individually predicted scattered fractions can be shown to remain  unchanged if all length parameters, i.e. object size, grid spacing and wavelength, are scaled by the same factor. 
For our comparison we have chosen $D/\lambda=10$ as is typical for single particle XUV scattering experiments~\cite{Rupp2012, Barke2015, Rupp2017, Langbehn2018}, such that the refractive index is the only remaining free parameter. Note that using a wavelength substantially smaller than the particle diameter ensures that the spatial resolution is sufficient to resolve detailed shape information. 
In our analysis we compare the diffraction images for scattering angles up to $\vartheta=45^{\circ}$ in order to be sensitive to wide-angle diffraction features.

\ifdefined\printImages
\begin{figure*}[h!]
    \fontsize{8}{10}\selectfont
    \centering
    \includegraphics[width=1\hsize]{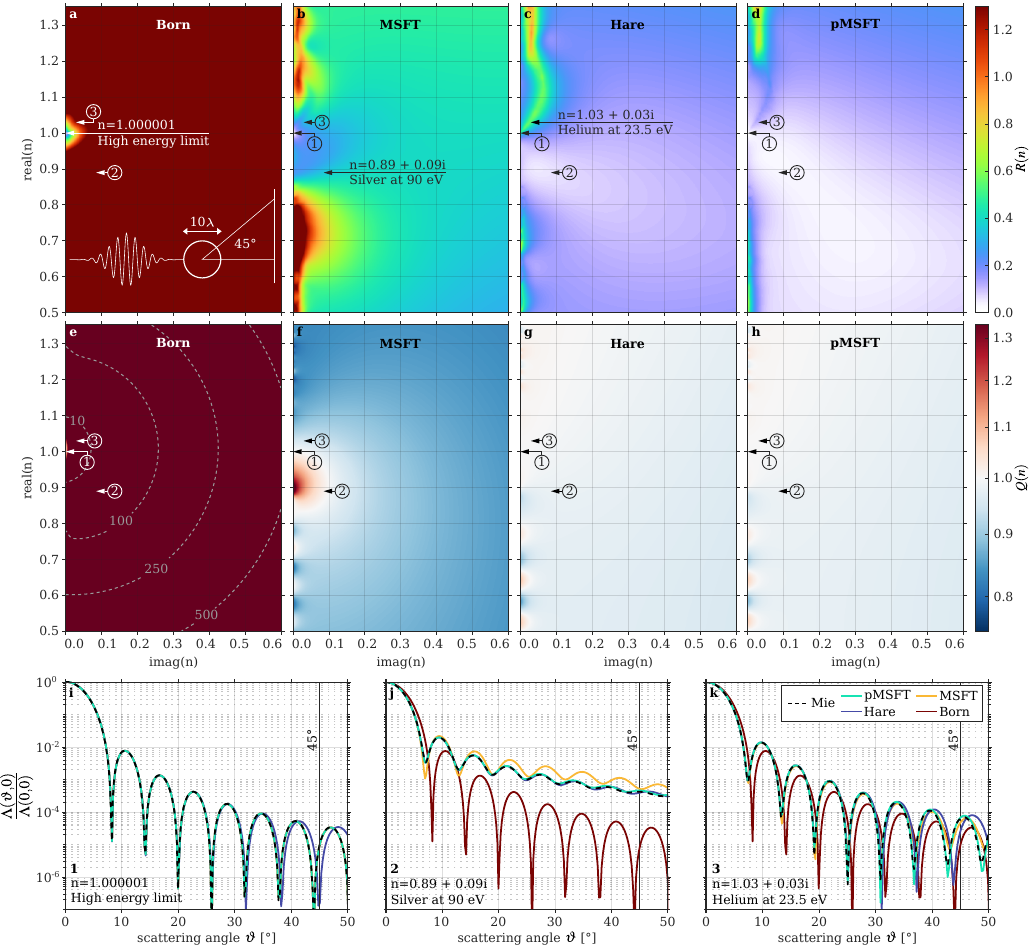}
    \caption{Systematic benchmark of the different simulation methods for the scattering off a homogeneous sphere (diameter $D=10\lambda$) as a function of the real and imaginary part of the refractive index.
    False-color representations in (a-d) and (e-h) display the feature error $R(n)$  and the relative signal strength $Q(n)$ when compared to a Mie reference calculation, respectively (see text). As the signal strength in Born's approximation (e) greatly exceeds the range of the shared colorbar essentially everywhere, we have added grey contour lines for selected error levels.  
    Representative angular profiles for three typical experimental scenarios, i.e. for the high energy limit with \mbox{$n_1=1.000001$} (i), for silver at 90 eV with $n_2=0.89 + 0.09$i (j), and for helium at 23.5 eV with $n_3=1.03 + 0.03$i (k) illustrate the angle-dependent deviation underlying the aggregated accuracy measures (a-h). The data in (i-k) reflects the normalized scattered fraction in the plane with normal vector parallel to the polarization direction ($\varphi=0$).}
    \label{fig:benchmark_results_feature}
\end{figure*}
\fi
The benchmark calculations have been evaluated regarding two different aspects, i.e. the correct prediction of relative signal features and absolute signal strength. In typical experiments the extraction of the essential shape information can be achieved based on the relative brightness and position of the  features in the scattering image, without knowledge of the absolute signal strength. A quantitatively correct description is required for the prediction of the absolute image brightness, the contrast of interferences structures resulting from regions with different scattering strength (e.g. for core-shell systems), or the quantitative retrieval of the optical properties. For analyzing these different aspects we employ the two quality measures
\begin{align}
    R(n)&= \int_{\vartheta\leq45^\circ} \biggr| \log \frac{\Lambda_{\mathrm{i}}(\vartheta, \varphi)}{\Lambda_{\mathrm{i}}(0,0)}- \log \frac{\Lambda_{\mathrm{ref}}(\vartheta, \varphi)}{\Lambda_{\text{ref}}(0,0)}\biggl|\,d\Omega
    \qquad \text{and}\\
    Q(n)&=\frac{\Lambda_i(\varphi=0, \vartheta=0)}{\Lambda_{\mathrm{ref}}(\varphi=0, \vartheta=0)}, 
\end{align}
where $R(n)$ quantifies the cumulative relative error of the scattered fraction, $Q(n)$ reflects the quantitative match of the absolute scattering signal in forward direction and the index i represents the respective method. Note, that the use of the logarithm in $R(n)$ ensures that errors for scattering angles with low and high signal contribute equally.

The analysis of both, the feature error via $R(n)$ and the absolute signal scaling via $Q(n)$ within the relevant refractive index range is displayed in \textbf{Figure~\ref{fig:benchmark_results_feature}}~\hyperref[fig:benchmark_results_feature]{(a-h)}.
In addition, normalized angular resolved scattering profiles are shown in Figure~\ref{fig:benchmark_results_feature}~\hyperref[fig:benchmark_results_feature]{(i-k)} for representative optical properties as realized in typical experiments.

The Born method predicts correct features only for a refractive index very close to unity as realized in the high photon energy limit (see Figure~\ref{fig:benchmark_results_feature}~\hyperref[fig:benchmark_results_feature]{(i)}) and deviates strongly otherwise (see Figure~\ref{fig:benchmark_results_feature}~\hyperref[fig:benchmark_results_feature]{(j,k)}. When employed in the XUV or soft X-ray range, the Born method can therefore only deliver a rough estimate of the scattering features. Furthermore it substantially overestimates the scattering signal in the entire refractive index range analyzed and is thus not suitable for quantitative predictions (see Figure~\ref{fig:benchmark_results_feature}~\hyperref[fig:benchmark_results_feature]{(e)}).
Note that a systematic comparison with the SAXS approach has been omitted, as it performs, at best, similarly as the Born method. 
The MSFT method reflects a substantial advance over Born and provides a reasonable prediction of features in most regions of the relevant optical property range (cf. Figure~\ref{fig:benchmark_results_feature}~\hyperref[fig:benchmark_results_feature]{(j,k)}). Figure~\ref{fig:benchmark_results_feature}~\hyperref[fig:benchmark_results_feature]{(b)} shows that accurate results are obtained in the vicinity of $n=1$ while strong deviations occur for $\mid\mathcal{R}(n)-1\mid\geq0.1$ in combination with very small imaginary parts $\mathcal{I}(n)\leq0.1$. Regarding the absolute signal strength, MSFT allows reasonable estimates with deviations of up to ca. $\pm25\%$ for the considered scenarios (cf. Figure~\ref{fig:benchmark_results_feature}~\hyperref[fig:benchmark_results_feature]{(f)}).

The essential step of improvement is achieved with the two split-step methods as they enable an exquisite predictions of features in a wide range of optical properties. They further allow accurate predictions of the signal strength with deviations not exceeding a few percent (cf. Figure~\ref{fig:benchmark_results_feature}~\hyperref[fig:benchmark_results_feature]{(g,h)}).
A direct comparison shows that the additional paraxial approximation implemented in Hare's split step method leads to a considerable deviation with errors even exceeding that of the MSFT prediction in the vicinity of $n=1$ (cf. Figure~\ref{fig:benchmark_results_feature}~\hyperref[fig:benchmark_results_feature]{(i)}) and a minute increase of the feature error in the entire analyzed refractive index range. 
Note that the significant deviation of Hare's prediction for scattering angles $\vartheta \geq 45^\circ$ in Figure~\ref{fig:fdtd_vs_pmsft}~\hyperref[fig:fdtd_vs_pmsft]{(b,d)} demonstrates the breakdown of the paraxial approximation for the description of large angle scattering.
Since the numerical effort for both split step approaches is identical, pMSFT should be the preferred method.
\section{Conclusion}
\label{sec:conclusion}
We have introduced the propagation multi-slice Fourier transform method as an accurate and numerically tractable method for the approximate description of light scattering off arbitrarily shaped targets.
Our rigorous derivation based on first principles reveals a unified picture of the physics included in pMSFT and in all other common methods used in the field such as SAXS, Born, MSFT, and Hare's split step method. 
We present a notation that enables quantitative predictions with all of these methods.
We have benchmarked the results of all methods against FDTD and Mie reference calculations to highlight their respective feature reproduction capabilities and ranges of applicability.
The superior accuracy of pMSFT combined with it's numerical efficiency makes it the de-facto method of choice for wide-angle scattering in the XUV and soft X-ray range.

The central conclusion from our study is that ultimately only the SAXS and pMSFT methods need to be considered for practical applications. The SAXS approach is the only method with a better numerical scaling than pMSFT, at the cost of very strong restrictions regarding applicability. In particular, the SAXS method is only justified when the refractive index of the target is close to unity and only small angles are relevant, as often realized in the X-ray regime.
In all other cases, the pMSFT approach should be the method of choice.

\medskip
\noindent
\textbf{Acknowledgements} \par 
\noindent
We thank Björn Kruse for fruitful discussions and support regarding the analysis of the FDTD results. We acknowledge financial support from  the German Research Foundation (DFG) via CRC 1477 ”Light-Matter Interactions at Interfaces” (ID: 441234705) and from the
European Social Fund and the Ministry of Education, Science and Culture of Mecklenburg-Vorpommern via project NEISS (ID: ESF/14-BM-A55-0007/19).

\medskip

\bibliographystyle{MSP}
\bibliography{diffraction_citations}

\end{document}


\newcommand{\wavtrans}{T}
\def\printImages{}  

\renewcommand{\rm}[1]{\text{#1}}

\title{Supporing Information on: Fast Simulation of Wide-Angle Coherent Diffractive Imaging}
\date{}
\maketitle
P. Tuemmler, J. Apportin, T. Fennel*, C. Peltz*\\
University of Rostock, Institute of Physics\\
Albert-Einstein-Str. 23 18059 Rostock, Germany\\
thomas.fennel@uni-rostock.de, christian.peltz@uni-rostock.de


\section{Propagation equation}\label{section:apdx_prop_eq}
We depart from the medium Maxwell  equations in coordinate space and time domain (e.g. $\vec E= \vec E (\vec r,t)$) for unmagnetic media ($\vec B =\mu_0 \vec H$) and without external charges or currents
\begin{align}
\mathrm{div} \,\vec D&=0 &\qquad \quad  \mathrm{div}\, \vec B&=0 \\ 
    \mathrm{rot}\,  {\vec E}&=-\dot {\vec  B} &\qquad  \mathrm{rot}\,  {\vec B}&=\mu_0 \dot {\vec D}. 
\end{align}
The displacement field is defined via
\begin{align}    
    \qquad \vec D=\varepsilon_0 \vec E+ \vec P, 
    \label{eq_def_displacement}
\end{align}
where $P$ is the polarization. Combining the curl equations and using $\mathrm{rot}\,\mathrm{rot}=(\nabla \nabla \cdot) - \Delta$ yields
\begin{align}    
    \nabla \nabla \cdot \vec E - \Delta \vec E=-\frac{1}{c_{0}^2} \ddot {\vec E}- \mu_0  \ddot {\vec P},
    \label{eq_wave1}
\end{align}
with the vacuum speed of light $c_0=1/\sqrt{\mu_0 \varepsilon_0}$. We now transform into the frequency domain using the Fourier transforms
\begin{align}
    f(t)&=\frac{1}{\sqrt{2\pi}}\int f(\omega) e^{-i\omega t}\, dt\qquad \mbox{and}\\
    f(\omega)&=\frac{1}{\sqrt{2\pi}}\int f(t) e^{i\omega t}\, dt
\end{align} 
As a result, time derivatives turn into factors ($\frac{\partial}{\partial t}\rightarrow -i\omega$) such that the scalar wave equation in the spectral domain reads 
\begin{align}    
    \left(-(\nabla \nabla \cdot) +\Delta + \frac{\omega^2}{c_0^2} \right)\vec E(\vec r,\omega )= -\mu_0 \omega^2 {\vec P} (\vec r,\omega ).
\end{align}
Combining Equation~\eqref{eq_def_displacement} and Equation~\eqref{eq_def_displacement}, the divergence contained in the first term on the left hand side can be expressed as 
\begin{align}    
\nabla \cdot  {\vec E}(\vec r,\omega)=-\frac{1}{\varepsilon_0}\nabla \cdot \vec P(\vec r,\omega) \label{eq:wave_equation_appendix}
\end{align}
For an isotropic linear medium the polarization can be described as   
\begin{align}    
\vec P(\vec r,\omega)=\varepsilon_0 \chi(\vec r,\omega) \vec E (\vec r,\omega)
\end{align}
As a result, the divergence term becomes
\begin{align}    
\nabla \cdot \vec E(\vec r, \omega)=- \frac{\nabla \chi(\vec r,\omega)}{1+\chi(\vec r,\omega)} \cdot \vec E(\vec r, \omega). \label{eq:neglecting_divergence}
\end{align}
Neglecting this term substantially simplifies wave equation~\eqref{eq:wave_equation_appendix}, as vectorial field components become uncoupled. The resulting scalar wave equation for angular frequency $\omega$
\begin{align}    
\left[ \Delta + k_0^2(1+\chi(\vec r, \omega))\right ] E(\vec r, \omega)= 0
\end{align}
with wave number $k_0=\frac{\omega}{c_0}$, applies to each field component separately and even yields exact results for propagation through homogeneous media as well as through interfaces under normal incidence. The high quality of the predictions using the scalar wave equation justify its applicability for our scenarios (cf. Section 5 in the main text).

\section{Polarization effect}\label{section:polarization_effect}
In the scalar description of the field propagation achieved by the neglect of the term described in Equation~\eqref{eq:neglecting_divergence}, polarization effects are not fully accounted for. In particular, the orthogonality of the electric field and the wave vector for propagating waves cannot be enforced. As a result, the signal loss for scattering along the direction of the incident electric field is absent. This polarization effect can be effectively reintroduced by a direction dependent scaling coefficient determined within single scattering approximation. 

\begin{figure}[!h]
    \centering
    \includegraphics[width=1\hsize]{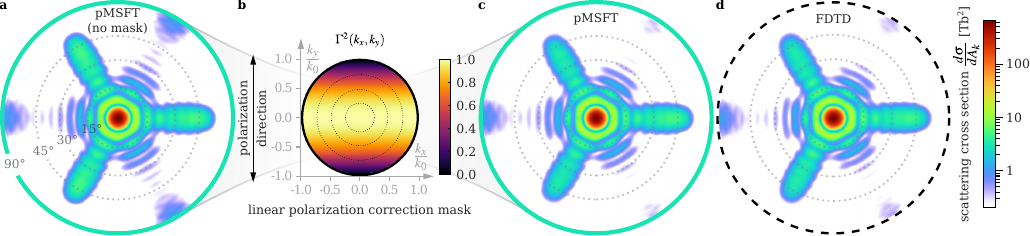}
    \caption{Impact of the polarization correction for the scenario from Figure 1 in the main text.
    Multiplication of the uncorrected diffraction pattern from pMSFT (a) with the correction mask $\Gamma^2(k_x, k_y)$ (b) yields the polarization-corrected pMSFT result (c).
    The improvement in the quantitative prediction of scattering features at large angles is illustrated by comparison to the corresponding FDTD result (d).}
	\label{fig:pol_mask}
\end{figure}

To this end, an incident electric field with linear polarization along $\vec e_y$ and propagating along $\vec e_z$ is considered to scatter at a point-like object. The polarization factor for scattering in direction $\vec k$ follows from the solution for a Hertzian dipole as
\begin{align}
    \Gamma(\vec k)=\vec e_k \times (\vec e_k \times \vec e_y) \quad \text{with} \quad \vec e_k = \vec k/k_0.
\end{align}
The evaluation of the two vector products yields
\begin{equation}
\Gamma(k_x, k_y)=\sqrt{1-\frac{k_y^2}{k_0^2}}.
\end{equation}
An example for the effect of the polarization correction is shown in \textbf{Figure~\ref{fig:pol_mask}} and highlights its importance for the description of scattering to large angles.

\medskip
